\documentclass{article}
\usepackage{ijcai15}

\usepackage{courier}
\usepackage{multirow}
\usepackage{float}
\usepackage{latexsym}
\usepackage{amssymb}
\usepackage{amsmath}
\usepackage{subfigure}
\usepackage{graphicx}
\usepackage{multicol}

\usepackage{algorithm}
\usepackage{algorithmic}

\usepackage{times}
\title{Robust Learning for Repeated Stochastic Games via Meta-Gaming}
\author{Jacob W. Crandall \\
Masdar Institute of Science and Technology\\
Abu Dhabi, UAE \\
jcrandall@masdar.ac.ae}

\begin{document}

\maketitle

\begin{abstract}
In repeated stochastic games (RSGs), an agent must quickly adapt to the behavior of previously unknown associates, who may themselves be learning.  This machine-learning problem is particularly challenging due, in part, to the presence of multiple (even infinite) equilibria and inherently large strategy spaces.  In this paper, we introduce a method to reduce the strategy space of two-player general-sum RSGs to a handful of expert strategies.  This process, called {\sc mega}, effectually reduces an RSG to a bandit problem.  We show that the resulting strategy space preserves several important properties of the original RSG, thus enabling a learner to produce robust strategies within a reasonably small number of interactions.  To better establish strengths and weaknesses of this approach, we empirically evaluate the resulting learning system against other algorithms in three different RSGs.
\end{abstract}

\section{Introduction}
In repeated stochastic games (RSGs), an agent must learn robust strategies within a handful of interactions when associating with other (unknown) agents who may also be learning.  This learning problem is challenging for several reasons.  First, strategy spaces of RSGs are inherently large, even for simple scenarios.  Second, the strategies used by other agents are unknown and can change over time, which produces a non-stationary environment.  Finally, the existence of multiple (even infinite) equilibria in general-sum RSGs renders rationality assumptions and equilibrium computation insufficient.   As a result of these challenges, existing learning algorithms (e.g., \cite{ClausBoutilier,minimaxQ,HuandWellman,friendorfoe,WoLF,CEQ,Pepper}) often fail to learn robust strategies within realistic time scales.

Game abstraction \cite{Gilpin2006,Schnizlein2009,Ganzfried2012,SandholmSingh2012} has emerged in recent years to address the first of these challenges.  Typically, the game is first reduced to a smaller game.  Next, an equilibrium strategy is computed for this smaller game, which is then executed in the original game.  This approach is effective in large zero-sum RSGs due to the unique nature of equilibria in these games.  Unfortunately, the usefulness of this approach is limited, since it is unclear in general-sum RSGs which equilibrium should be computed.

In this paper, we analyze an alternative form of strategy reduction for two-player general-sum RSGs.  Our approach, called {\sc mega}, reduces an RSG to a multi-armed bandit problem by computing a finite set of expert strategies (based on equilibria computations and learning rules).  This greatly simplifies the learning problem such that simple expert algorithms, such as Exp3~\cite{auer95gambling}, UCB~\cite{UCB}, EEE~\cite{exextrade}, or S++~\cite{CrandallJAIR2014}, can be used to learn on the reduced strategy space.

We make three contributions.  First, we propose {\sc mega} as a potentially suitable method for learning in two-player general-sum RSGs.  Second, we show that the strategy space computed by {\sc mega} preserves several important theoretical properties of the original RSG.  Finally, to better establish strengths and weaknesses of this approach, we empirically evaluate the robustness of the resulting learning system against other algorithms in three RSGs.

\section{Repeated Stochastic Games}
We first formally define and motivate RSGs.

\subsection{Notation}
We consider two-player RSGs played by players $i$ and $-i$.  An RSG consists of a set of {\em stage games} $S$.  In each stage $s \in S$, both players choose an action from a finite set.  Let $A(s) = A_i(s) \times A_{-i}(s)$ be the set of {\em joint actions} available in $s$, where $A_i(s)$ and $A_{-i}(s)$ are the action sets of players $i$ and $-i$, respectively.  Each {\em round} of an RSG begins in the start stage $\hat{s} \in S$ and terminates when some goal stage $s_g \in G \subseteq S$ is reached.  A new round then begins in stage $\hat{s}$.  The game repeats for an unknown number of rounds. 

When joint action $\mathbf{a} = (a_i,a_{-i})$ is played in $s$, the players receive the finite rewards $r_i(s,\mathbf{a})$ and $r_{-i}(s,\mathbf{a})$, respectively.  The world also transitions to some new stage $s'$ with probability defined by $P_M(s,\mathbf{a},s')$.  We assume that $P_M$, $r_i(s,\mathbf{a})$, and $r_{-i}(s,\mathbf{a})$ are known by both players {\em a priori}, and that the players can observe $s$ and each other's actions.  

Player $i$'s strategy, denoted $\pi_i$, defines how it will act in each world {\em state}.  In general-sum RSGs, it is often useful (and necessary) to define state not only in terms of the stage, but also in terms of the history of the players' actions.  Let $H$ denote the set of possible joint-action histories.  Then, the set of states is given by $\Sigma = S \times H$.  Let $\pi_i(\sigma)$ denote the policy of player $i$ in state $\sigma = (s, h) \in \Sigma$.  That is, $\pi_i(\sigma)$ is a probability distribution over the action set $A_i(s)$.

\subsection{Metrics}
The success of an algorithm in an RSG is measured by the payoffs it receives.  An ideal algorithm will maximize its payoffs against any associate.  Because this is a difficult task to achieve, measure, and guarantee, previous work has focused on identifying algorithms that meet certain criteria, such as convergence to Nash equilibria (NEs) \cite{HuandWellman,friendorfoe,WoLF}, Pareto optimality \cite{PowersIJCAI2005}, and security \cite{FictitiousPlay,PowersIJCAI2005}.

In this paper, we adopt a different, though related, metric of success.  We evaluate an algorithm based on the proportion of a population that is willing to use it, as determined by evolutionary simulations \cite{ReplicatorDynamic}.  Success in such simulations requires an algorithm to demonstrate many of the previously mentioned attributes.

\begin{figure}
\begin{center}
\subfigure[Player 1's tasks]{\label{fig:tasks1}
{\scriptsize
\setlength{\tabcolsep}{3pt}
\begin{tabular}[b]{cccc} \hline
{\bf Task} & {\bf Time} & {\bf Load} & \multirow{2}{*}{\bf Utility} \\
{\bf ID} & {\bf window} & {\bf (units)} &  \\ \hline
1 & [0,8) & 2.0 & 7.0 \\
2 & [5,8) & 2.0 & 1.5 \\
3 & [8,12) & 3.6 & 0.8 \\
4 & [10,11) & 2.4 & 1.6 \\
5 & [11,13) & 3.9 & 2.7 \\
6 & [14,17) & 3.8 & 1.4 \\
7 & [17,18) & 3.6 & 2.9 \\
8 & [18,21) & 1.2 & 1.5 \\
9 & [18,23) & 1.5 & 2.4 \\
10 & [23,24) & 5.0 & 20.2 \\ \hline 
\end{tabular}}
}~~~
\subfigure[Player 2's tasks]{\label{fig:tasks2}
{\scriptsize
\setlength{\tabcolsep}{3pt}
\begin{tabular}[b]{cccc} \hline
{\bf Task} & {\bf Time} & {\bf Load} & \multirow{2}{*}{\bf Utility} \\
{\bf ID} & {\bf window} & {\bf (units)} &  \\ \hline
11 & [0,3) & 1.5 & 2.0 \\
12 & [4,6) & 5.0 & 22.2 \\
13 & [7,8) & 1.5 & 0.9 \\
14 & [9,13) & 1.3 & 1.4 \\
15 & [11,15) & 0.7 & 2.4 \\
16 & [13,17) & 4.5 & 2.6 \\
17 & [15,18) & 2.7 & 1.7 \\
18 & [17,18) & 5.0 & 1.6 \\
19 & [18,22) & 2.8 & 1.5 \\
20 & [22,23) & 4.0 & 5.7 \\ \hline 
\end{tabular}}
}\\
\subfigure[Electricity generated per hour]{\label{fig:generation}\includegraphics[width=1.7in]{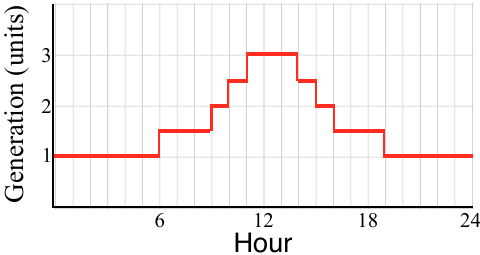}~~~}
\subfigure[Performance (self play)]{\label{fig:self}  \includegraphics[width=1.5in]{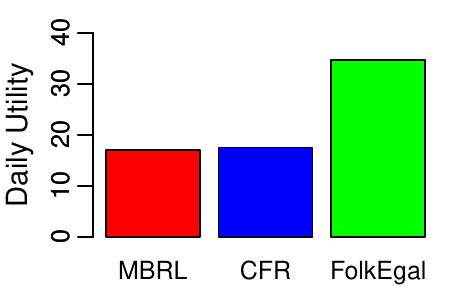}}
\end{center} 
\caption{The Microgrid Scenario.}
\label{fig:microgridwar}
\end{figure}

\subsection{Illustrative Example: A Microgrid Scenario}
Consider a microgrid in which two players share the limited electricity supply with per-hour generation characteristics shown in Figure~\ref{fig:microgridwar}c.  The players can store up to five units of unused electricity, though unused storage is lost at the end of the day.  To gain utility, a player executes its tasks, each requiring the specified electricity loads, within the stated time windows (Figure~\ref{fig:microgridwar}a-b).  A task is completed within a single hour and can be executed no more than once a day.  If the players try to consume more electricity than is available, a blackout occurs.  In a blackout, the electricity storage empties, and the tasks that the players attempted to execute are left unfinished.  A cost of two utility units is incurred by each player that attempted to execute tasks in that hour.

A stage is defined by the current hour, the amount of stored electricity, and the set of current tasks (unexecuted tasks whose time windows correspond to the current hour).  The game has 2,033 unique stages.  The start stage occurs in hour 0.  The goal stages are all stages with hour 24.  Each player's action set is the power set of its set of current tasks.

Insufficient electricity is generated each day for all tasks to be executed.  Furthermore, some high-valued tasks can only be executed when the other player complies.  For example, there is only sufficient electricity for Player 2 to execute Task 12 if Player 1 refrains from executing Task 1 prior to hour 6.  Similarly, Player 1 can only successfully execute Task 10 if Player 2 refrains from executing Task 20.  Thus, the players must coordinate and (likely) compromise to be successful.

To enforce a compromise, a player may need to punish its associate for undesirable behavior.  For example, to deter player 2 from executing Task 20 in the future, player 1 should punish player 2 whenever player 2 executes Task 20 (perhaps by executing Task 1 prior to hour 6 the next day so that player 2 cannot execute Task 12).  However, this strategy will only be successful if both players use a state space that includes at least some aspects of $H$.

However, using $H$ as a component of state makes a player's strategy space massive.  Even if limited to pure strategies and if $|A_i(s)| = 2$ for all $s \in S$, the size of $i$'s strategy space is on the order of $2^{| S |{| H |}}$.  Since only a small fraction of strategies can be executed during most interactions, traditional learning algorithms that operate on the full strategy space are unlikely to be successful.  Furthermore, it is unclear how traditional methods for game abstraction can be used since this RSG has an infinite number of NEs \cite{GintisGameTheory}.

Figure~\ref{fig:microgridwar}d compares three different algorithms in the Microgrid Scenario in self play, two of which are model-based reinforcement learning (MBRL) (see Appendix~\ref{app:mbrl} for details) and CFR \cite{Zinkevich2007,Johansonetal2012}.  Since $|S \times H|$ is prohibitively large, $H$ is discarded so that $\Sigma = S$.  Both algorithms converge to solutions in which player 1 does not execute Task 10 and player 2 does not execute Task 12.  Thus, both players achieve low utility.  

CFR and MBRL cannot learn cooperative behavior in this game because their representations (when discarding $H$) do not allow them to see the value in doing so.  Additionally, they do not consider punishment strategies that enforce cooperation.   On the other hand, because FolkEgal \cite{FolkEgal} computes and implements a trigger strategy focused on the egalitarian (cooperative) solution, it performs much better than MBRL and CFR in self play.  However, FolkEgal is ineffective when the associate does not play its portion of the egalitarian solution -- it does not learn.
	
\section{Meta-Gaming}

These latter results demonstrate how difficult it is to represent and learn non-myopic solutions in general-sum RSGs.  In this section, we describe how {\em me}ta-{\em ga}ming (in the form of a new algorithm called {\sc mega}) can be used to reduce the strategy space of the game so that a player can both model non-myopic (cooperative) solutions and adapt to its associate's behavior.

{\sc mega} (Algorithm~\ref{alg:mega}) reduces the strategy space of an RSG to a finite set $\Phi_i$ of strategies or algorithms.  Each $\phi \in \Phi_i$ defines a policy for each state $s \in S$.  Thus, {\sc mega} reduces an RSG to a multi-armed bandit problem, where each {\em arm} is an expert $\phi \in \Phi_i$.  Thus, rather than learning a separate policy in each state $s \in S$, the agent must learn which high-level strategies or algorithms $\phi \in \Phi$ are the most profitable.

\begin{algorithm}[tb]
   \caption{{\sc mega}}
   \label{alg:mega}
{\small
\begin{algorithmic}
    \STATE {\bfseries Input:} An expert algorithm ${\mathcal A}$
   \STATE {\bf Initialize:} Compute a set $\Phi_i$ of experts
   \STATE {\bf Run:} In each round $t$
   \STATE ~~~~~~~~~ - Select an expert $\phi_t \in \Phi_i$ using ${\mathcal A}$
   \STATE ~~~~~~~~~ - Follow the strategy prescribed by $\phi_t$ throughout round $t$
   \STATE ~~~~~~~~~ - Update each $\phi \in \Phi_i$ and ${\mathcal A}$ as specified
\end{algorithmic}}
\end{algorithm}

{\sc mega} has similarities to a method proposed by Elidrisi et al.~(\citeyear{Elidrisi2014}), in which meta-analysis is used to reduced an RSG to a normal-form game.  This method relies on exploration strategies, and clustering and thresholding algorithms to identify common sequences of actions (or paths) for each player.  {\sc mega}, instead, uses a variety of game-theoretic valuations (among other techniques) to define the set $\Phi_i$.  These valuations can be used in any RSG of any complexity.  Regardless of the game's complexity, {\sc mega} produces a handful of high-level strategies over which an expert algorithm learns.


{\sc mega} makes the learning problem for RSGs similar in nature to ensemble methods \cite{EnsembleMethods}, which have been used successfully in the machine learning and AI literature.  However, we are unaware of previous work that uses such methods for learning in general-sum RSGs.  Determining how to effectively reduce the strategy space of an arbitrary RSG to a small set of strategies is non-trivial.

To implement Algorithm~\ref{alg:mega}, we must solve two technical problems.  First, we must identify an expert algorithm ${\mathcal A}$ that learns effectively in repeated interactions with unknown associates.  This problem has been well-studied in the literature \cite{auer95gambling,GIGA-WoLF,Arora2012,CrandallJAIR2014,CesaBianchi2013}.  Thus, we focus on the second technical problem: defining an effective set of experts $\Phi_i$.  Ideally, in any scenario that a player is likely to encounter, at least one expert $\phi \in \Phi_i$ should perform well.  However, no single expert need be effective in all scenarios.   

\section{Experts}
\label{sec:experts}

Our set $\Phi_i$ consists of three types of experts: leader strategies, follower strategies, and preventative strategies.  We introduce {\em preventative strategies} in this paper.  The idea of leader and follower strategies was first identified by Littman and Stone (\citeyear{LittmanLeaderAlgs}) in the context of repeated normal-form games.  We define a set of such experts for RSGs. 

\subsection{Leader Strategies}
A leader strategy encourages its associate to follow a {\em target solution}, by playing its own portion of the target solution as long as its associate plays its part.  When the associate deviates from the solution, the leader subsequently retaliates so that the associate does not profit from the deviation.

We define our leader experts by (1) identifying potential target solutions, (2) selecting a subset of these solutions, one corresponding to each leader in $\Phi_i$, and (3) defining a punishment (retaliatory) strategy.

\subsubsection{Computing Target Solutions}
\label{sec:compute}
Possible target solutions are computed by solving Markov decision processes (MDPs) over the joint-action space of the RSG.  The MDPs are defined by $P_M$, the joint-action set $A$, and a payoff function that is a convex combination of the players' rewards \cite{FolkEgal}.  That is, for $\omega \in [0,1]$, the payoff function is $y^\omega(s,\mathbf{a}) = \omega r_i(s,\mathbf{a}) + (1-\omega) r_{-i}(s,\mathbf{a})$.  Then, the value of joint-action $\mathbf{a}$ in state $s$ is
\begin{eqnarray}
Q^\omega(s,\mathbf{a}) = y^\omega(s,\mathbf{a}) + \sum_{s' \in S} P_M(s,\mathbf{a},s') V^\omega(s'),
\end{eqnarray}
where $V^\omega(s) = \max_{\mathbf{a} \in A(s)} Q^\omega(s,\mathbf{a})$.  The MDP can be solved in polynomial time using linear programming \cite{PapadimitriouTsitsiklis1987,LittmanMDP}.  

Let MDP($\omega$) denote the joint strategy produced by solving an MDP for a particular $\omega$.  Also, let $V_i^\omega(s)$ be player $i$'s expected future payoff from stage $s$ when MDP($\omega$) is followed.  Then, the ordered pair $\left(V_i^\omega(\hat{s}), V_{-i}^\omega(\hat{s}) \right)$ is the joint payoff vector for the target solution defined by MDP($\omega$).  This payoff vector is Pareto optimal \cite{FolkEgal}.

By varying $\omega$, we can compute a variety of possible target solutions (called {\em pure solutions}).  Additional possible target solutions, or {\em alternating solutions}, are obtained by alternating between different pure solutions.  For example, in the Microgrid Scenario, MDP(0.1) and MDP(0.3) produce the joint payoffs $(11.3, 40.0)$ and $(36.8, 32.7)$, respectively.  Alternating between these solutions produces the average joint payoff $(24.05, 36.35)$.  Since longer cycles are difficult for associates to model, we only include cycles of length two.

\begin{figure}[t]
\begin{center}
\includegraphics[width=2.3in]{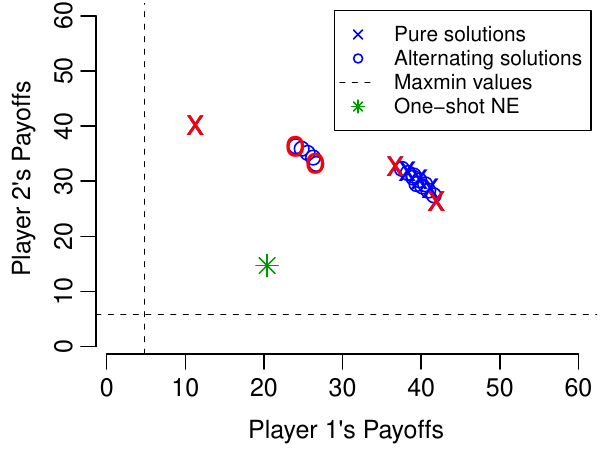}
\end{center}
\caption{Joint payoffs of possible target solutions in the Microgrid Scenario.  Red indicates selection by our method.}
\label{fig:solutions} 
\end{figure}

\subsubsection{Which Target Solutions?}
\label{sec:which}
Figure~\ref{fig:solutions} shows the joint payoffs of possible pure and alternating target solutions in the Microgrid Scenario.  In this RSG, the one-shot NE is Pareto dominated by many possible target solutions.  Since any solution in which each player's payoff exceeds its maximin value can be sustained as NEs of the repeated game \cite{GintisGameTheory}, these possible target solutions offer a variety of potentially desirable equilibrium solutions.

A larger $\Phi_i$ provides more resolution, but makes the learning problem more difficult.  In this work, we form leader experts for up to five solutions.  We select the egalitarian solution and the two solutions that give each player its highest payoff subject to the other player getting its security level.  We then select two points that maximize the Euclidean distance from the other selected solutions.  This provides a variety of different options for the players to agree to.  Figure~\ref{fig:solutions} shows the selected points for the Microgrid Scenario in red.

\subsubsection{Adding punishment}
A typical leader punishes a deviation from the target solution by playing an attack strategy (usually its minimax strategy).  In practice, we deviate slightly in this regard: player $i$ only punishes deviations by $-i$ that substantially lower $i$'s payoffs.  Let $s_\tau$ be the $\tau$th stage of the round.  Player $i$ begins punishing $-i$ when $-i$ deviates from the target solution and
\begin{eqnarray}
\mbox{Condition~1:}~~V_i^{\omega}(s_{\tau}) + r_i^t({\tau-1}) <  V_i^{\omega}(s_{\tau-1}), \\
\mbox{Condition~2:}~~\sum_{j=1}^\tau r_i^t(j) + V_i^{\omega}(s_{\tau}) < \alpha_i^t. \mbox{\scriptsize ~~~~~~~~~~~~~~~~~~~}
\label{eq:conditions}
\end{eqnarray}
Here, $r_i^t(j)$ is $i$'s payoff after the $j$th move of round $t$, $\alpha_i^t = \lambda \alpha_i^{t-1} + (1-\lambda) R_i^t$, $\lambda \in (0,1)$, and $R_i^t$ is $i$'s total payoff in round $t$.  $\alpha_i^0$ is set to $i$'s payoff in the egalitarian solution.  In words, these two conditions identify when deviations by the associate have lowered player $i$'s expected round payoffs sufficiently to justify retaliation.

To punish player $-i$'s harmful deviations, player $i$ plays its minimax strategy for the remainder of round $t$, and continues to do so in subsequent rounds until $-i$'s payoffs are at least $\delta$ less than they would have been had $-i$ not deviated.  We adopt this punishment mechanism since the associate's payoffs are often uncertain, and rewards and transition functions in RSGs can be non-deterministic.  More leniency can sometimes eliminate cycles of unnecessary punishment.



\subsection{Follower Strategies}
Followers seek to maximize their payoffs against the strategy they attribute to their associate.  We include followers in $\Phi_i$ that estimate their associate's strategy in three ways.  The first set of followers assume the associate plays a leader strategy.  Against such associates, a player maximizes its payoffs by following the corresponding target solution.  We form a separate follower strategy for each selected target solution, each of which follows the target solution unconditionally.

The set $\Phi_i$ also includes two other follower strategies: MBRL (Appendix~\ref{app:mbrl}) and the maximin expert $\phi_i^{mm}$, which plays a best response to an associate seeking to minimize its payoffs.  Formally, let $u_i(\pi_i,\pi_{-i})$ be the expected utility in a round to player $i$ when it follows strategy $\pi_i$ and player $-i$ follows strategy $\pi_{-i}$.  Then, the strategy followed by $\phi_i^{mm}$ is
 \begin{eqnarray}
\pi_i^{mm} = \arg \max_{\pi_i \in \Pi_i} \min_{\pi_{-i} \in \Pi_{-i}} u_i(\pi_i,\pi_{-i}).
 \label{eq:maxmin}
 \end{eqnarray}

\subsection{Preventative Strategies}
Due to their state representations, many algorithms have difficulty perceiving the punishment signals communicated by leaders.  In some cases, {\em preventative strategies} can be more effective.  Rather than punishing past deviations, preventative strategies seek to make deviations unprofitable in the first place by anticipating deviations the associate might make, and then acting to make these deviations unprofitable.

We include one preventative strategy in $\Phi_i$, which we refer to as {\sc Bouncer}.  {\sc Bouncer} seeks to minimize the difference between the players' payoffs, without regard for its own payoffs.  Formally,  Bouncer computes both $Q_i(s,\mathbf{a})$ and $Q_{-i}(s,\mathbf{a})$ using SARSA \cite{Sarsa1994}, where the players' strategies are estimated using the Fictitious-play assessment.  It then selects action
\begin{eqnarray}
a_i^*(s) = \min_{a_i \in A_i(s)} \sum_{a_{-i} \in A_{-i}(s)} \gamma_{-i}(s,a_{-i}) U(s,(a_i,a_{-i})),
\label{eq:bouncer_act}
\end{eqnarray}
where $U(s,\mathbf{a}) = |Q_i(s,\mathbf{a}) - Q_{-i}(s,\mathbf{a})|$.  

\section{Properties of the Strategy Reduction}
Good strategy reductions for RSGs should maintain important attributes of the original strategy space, such as NEs \cite{HuandWellman,friendorfoe}, Pareto optimality \cite{PowersIJCAI2005}, best response \cite{WoLF}, and security \cite{FictitiousPlay,PowersIJCAI2005}.   Maintaining these properties helps to ensure that an algorithm confined to the reduced strategy space can learn an effective strategy in the original game.  In this section, we show that the strategy set $\Phi_i$ maintains three important properties of the original RSG.

 \vspace{.07in}

\noindent {\bf Property 1} (Security) {\em The strategy set $\Phi_i$ has the same security level as the strategy set of the original RSG.}

\vspace{-.1in}~\\ {\em Proof:} Player $i$'s security level in the original RSG is its maximin value, $v_i^{mm} = \max_{\pi_i \in \Pi_i} \min_{\pi_{-i} \in \Pi_{-i}} u_i(\pi_i,\pi_{-i})$.  That is, player $i$ can guarantee itself an expected payoff of at least $v_i^{mm}$ per round if it plays $\pi_i^{mm}$ (Eq.~\ref{eq:maxmin}).  Given the reduced strategy set $\Phi_i$, however, player $i$'s security level is $\hat{v}_i^{mm} = \max_{\phi_ \in \Phi_i} \min_{\pi_{-i} \in \Pi_{-i}} u_i(\phi_i,\pi_{-i})$.  We know two things about $\Phi_i$.  First, since $\Phi_i \subseteq \Pi_i$, $\hat{v}_i^{mm} \leq v_i^{mm}$.  Second, since $\pi_i^{mm} = \phi_i^{mm} \in \Phi_i$, we know that $\hat{v}_i^{mm} \geq v_i^{mm}$.  These two statements are only both true when $\hat{v}_i^{mm} = v_i^{mm}$.  $\square$

 
\vspace{-.07in}~\\ \noindent {\bf Property 2} (Best response) {\em When the associate always follows a stationary strategy that is Markovian in $S$, the strategy set $\Phi_i$ eventually contains a best response with respect to the strategy set of the original RSG.}

 \vspace{-.05in}~\\ {\em Proof sketch:} Strategy $\pi_i^*$ is a {\em best response} to the associate's strategy $\pi_{-i}$ if $\forall \pi_i \in \Pi_i$, $u_i(\pi_i^*, \pi_{-i}) \geq u_i(\pi_i, \pi_{-i})$.  MBRL, an expert in $\Phi_i$, computes a best response with respect to its assessment of its associate's strategy.  When the associate's strategy is stationary and Markovian with respect to $S$, this assessment converges to the strategy used by the associate given sufficient exploration and observation.  Thus, MBRL eventually converges to a best response in this case.   $\square$

~\\ \noindent
{\bf Property 3} (Nash equilibria) {\em $\Phi_i$ contains strategies that correspond to Nash equilibria of the original RSG.}

\vspace{-.05in}~\\ {\em Proof sketch:} The leader strategies in $\Phi_i$ are trigger strategies in which the target solution gives both players at least their security levels.  An associate's best response to the trigger strategy is to play its portion of that strategy's target solution.  Thus, when the associate plays a trigger strategy with the same target solution, the result is a NE (each player is playing a best response to the other's strategy).  Thus, each leader strategy in $\Phi_i$ corresponds to player $i$'s strategy in a NE of the original RSG.   $\square$

Since MDP($\omega$) produces a solution that is (approximately) Pareto optimal \cite{FolkEgal}, each leader strategy in the set $\Phi_i$ with a pure target solution corresponds to a Pareto optimal NE of the original RSG.  Alternating target solutions involve alternations between Pareto optimal solutions, but are not themselves guaranteed to be Pareto optimal.

\section{Empirical Evaluations}
We seek to identify how to quickly learn robust strategies in general-sum RSGs played with arbitrary associates.  We now use empirical evaluations to determine how well {\sc mega} helps meet this goal.  To do this, we paired {\sc mega} with the expert algorithms Exp3 \cite{auer95gambling} and S++ \cite{CrandallJAIR2014} to form {\sc mega-Exp3} and {\sc mega-S++}.  


\begin{figure}[t]
\begin{center}
\includegraphics[width=0.67in]{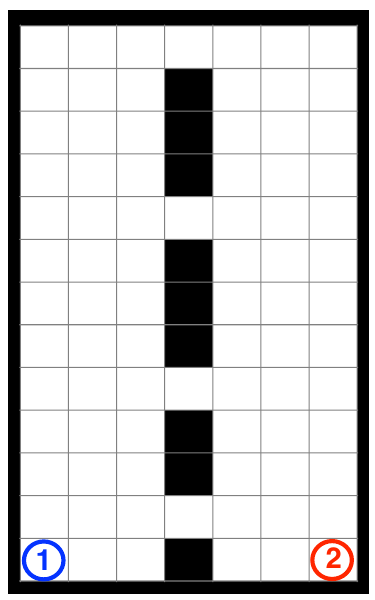} ~~~~~~~~~~~~~~~~
\includegraphics[width=1.08in]{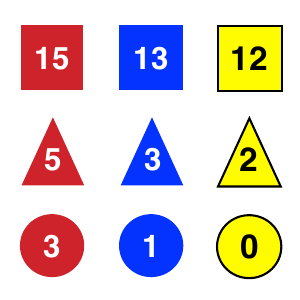}
\end{center} 
\caption{The SGPD (left) and a Block Game (right).}
\label{fig:moregames} 
\end{figure}

We evaluated {\sc mega-Exp3} and {\sc mega-S++} against ten algorithms in three RSGs: the Microgrid Scenario, the SGPD, and a Block Game.  The SGPD (a stochastic-game prisoners' dilemma) and the Block Game are described by Oudah {\em et al.} (\citeyear{Oudah2015}).  In the Block Game, two players share the block set shown in Figure~\ref{fig:moregames}b.  In each round, the players take turns selecting blocks until each player has three blocks.  If a player's blocks form a valid set (i.e., all blocks of the same color, all blocks of the same shape, or none of the blocks have the same color or shape), her payoff is the sum of the numbers on her blocks.  Otherwise, she loses the sum of the numbers divided by 4.  The sub-game perfect one-shot NEs of this game give each player 18 points.  However, these solutions are dominated by the solution in which the players alternate between taking all of the squares and all of the triangles (each player averages 25 points).  Even better, a player could potentially bully its associate by insisting it always gets all the squares.
  
\begin{figure}
\begin{center}
\setlength{\tabcolsep}{3pt}
{\scriptsize
\begin{tabular}[b]{lccc} \hline
	{\bf Algorithm} & {\bf Micro} & {\bf SGPD} & {\bf Block}\\ \hline
{\sc mega-S++} & 24.43 & 11.90 & 23.21 \\
{\sc CFR} & 21.19 & 9.75 & 20.83 \\
{\sc MBRL} & 20.95 & 8.52 & 19.46 \\
{\sc mega-Exp3} & 19.35 & 9.63 & 17.68 \\
{\sc FolkEgal} & 19.35 & 10.07 & 15.98 \\
{\sc u-MBRL} & 20.00 & 4.94 & 14.50 \\
{\sc Bully} & 12.08 & 9.30 & 17.60 \\
{\sc Friend} & 17.83 & 9.75 & 10.00 \\
{\sc Maxmin} & 17.29 & 9.75 & 5.77 \\
{\sc Bouncer} & 9.53 & 10.82 & 10.99 \\ \hline
& & & \\
& & & \\
\end{tabular}}
\includegraphics[width=1.7in]{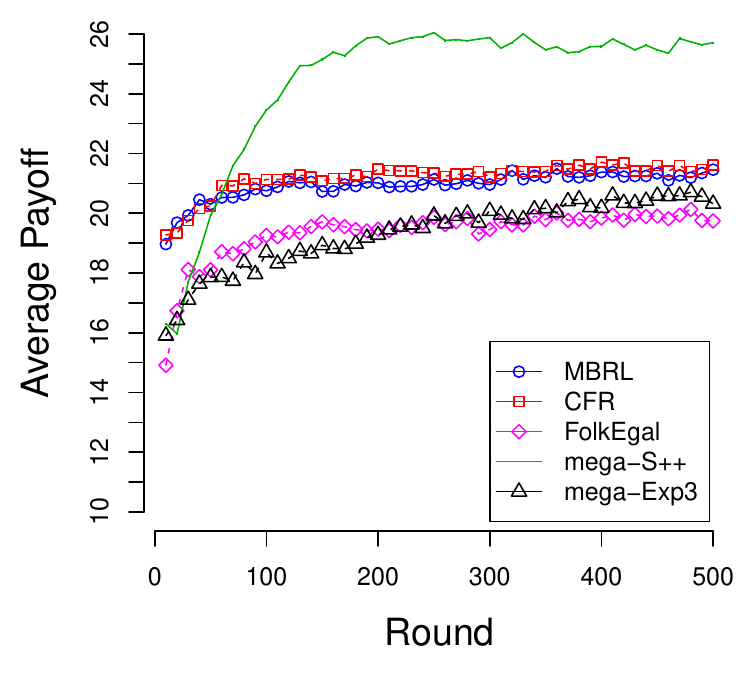}~
\end{center} 
\caption{Average payoffs (left) against all associates over 500 rounds, and (right) over time in the Microgrid Scenario.}
\label{fig:times_r}
\end{figure}

Each algorithm was paired with every other algorithm in each game.  The average of 25 trials was taken for each pairing.  In addition to the algorithms already mentioned, the ten algorithms (Figure~\ref{fig:times_r}a) included Friend-VI \cite{friendorfoe}, utilitarian MBRL (u-MBRL), and Bully.  u-MBRL is identical to MBRL except that it seeks to maximize the sum of the two player's payoffs (rather than just its own).  Bully is the leader expert that gives the agent its highest payoff.


\subsection{Performance}
The average per-round payoffs of each algorithm in each RSG, averaged over all pairings, are shown in Figure~\ref{fig:times_r}a.  While some of the algorithms varied in their effectiveness across games, a number of trends remained constant in each game.  First, in each RSG, {\sc mega-S++} had the highest average payoff in 500-round games.  CFR had the second highest average payoffs in the Microgrid Scenario and in the Block Game, though it finished fourth in the SGPD.  

Figure~\ref{fig:times_r}b shows the average payoffs of the agents over time in the Microgrid Scenario.  Results for the other two RSGs (not shown) are similar.  In each game, CFR had the highest average payoff over the first 50 rounds.  Thereafter, {\sc mega-S++} substantially outperformed CFR.  We have observed that CFR tends to produce rather myopic solutions in general-sum RSGs.  These solutions are relatively easy to establish, which leads to higher payoffs in early rounds.  On the other hand, the strategy reduction provided by {\sc mega} allows S++ to learn to establish cooperative solutions (when profitable) when associates are apt to cooperate.  These compromises are more difficult to establish, but have higher returns.

\begin{figure}
\begin{center}
\includegraphics[width=2.25in]{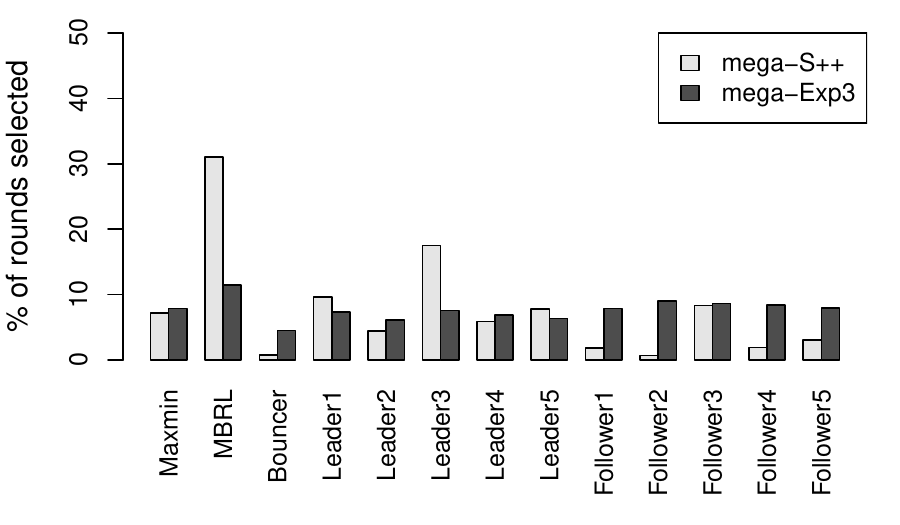}
\end{center} 
\caption{Percentage of rounds that each expert was selected by {\sc mega-S++} and {\sc mega-Exp3} in the Microgrid Scenario.}
\label{fig:selected}
\end{figure}

Though it learned on the same (reduced) strategy space, {\sc mega-Exp3} was not as successful as {\sc mega-S++}.  The reason for this is depicted in Figure~\ref{fig:selected}, which shows the percentage of rounds that each expert was selected by the two algorithms in the Microgrid Scenario.  {\sc mega-S++} rarely used some of the experts, while utilizing {\sc MBRL} and Leader3 (FolkEgal) extensively.  On the other hand, {\sc mega-Exp3} did not distinguish among the experts as much over the first 500 rounds, and hence failed to learn effective strategies.

\subsection{Evolutionary Robustness}
While {\sc mega-S++} achieved higher average payoffs in games lasting more than 50 rounds, this does not necessarily mean that agents will want to use it.  To help make this determination, we simulated populations of agents using the replicator dynamic \cite{ReplicatorDynamic} over 100 generations, and plotted the percentage of agents in the population that used each algorithm.  To be successful in such simulations, an algorithm must be able to avoid being invaded and must be able to associate effectively with other successful algorithms.

\begin{figure}
\begin{center}
~\\
\hspace{-.15in}
\subfigure[Microgrid Scenario]{\label{fig:gridwar1_usage}\includegraphics[width=1.7in]{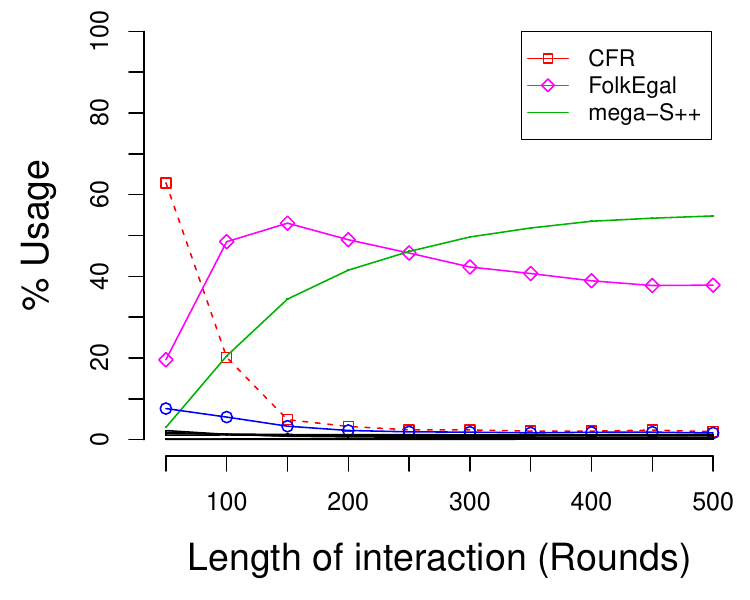}}
\subfigure[SGPD]{\label{fig:sgpd_usage}\includegraphics[width=1.7in]{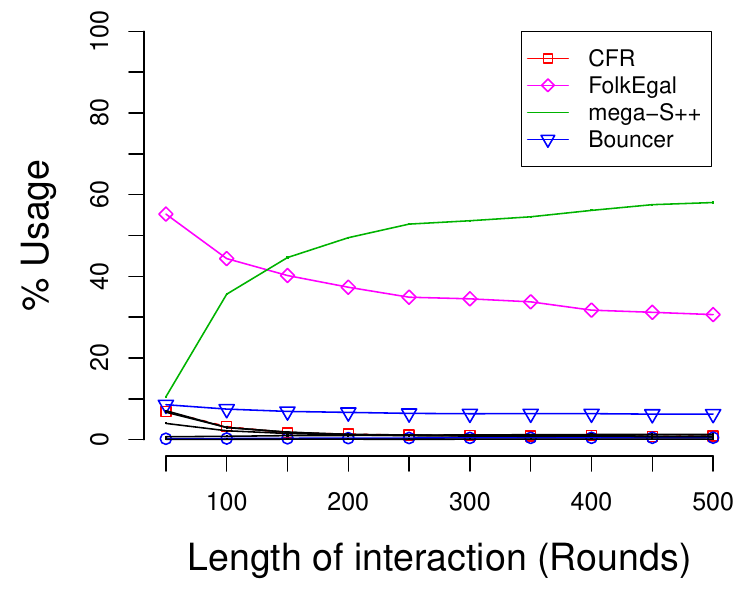}}
\end{center} 
\caption{Usage of each algorithm in evolutionary simulations over 100 generations  as a function of interaction length.}
\label{fig:usage}
\end{figure}

Figure~\ref{fig:usage} plots the average usage of each algorithm in the Microgrid Scenario and the SGPD as a function of the length of the interaction  (results are similar in the Block Game).  When agents interacted for shorter periods of time, FolkEgal and CFR (in the Microgrid Scenario and the Block Game) were used the most.  However, when interactions lasted longer, {\sc mega-S++} was used the most in each RSG.

These latter results are interesting for several reasons.  First, they show the robustness of {\sc mega-S++}, particularly in long-term interactions.  Second, while FolkEgal did not perform exceptionally well over all pairings, it was used extensively by the agents in the population.  Third, while CFR had high per-round payoffs averaged over all pairings, it was not used extensively in long-term interactions.  Table~\ref{tab:breakdown} provides an explanation of these phenomena.  The table shows that both {\sc mega-S++} and FolkEgal were effective in self play, whereas CFR was not.  Additionally, CFR tended to perform poorly against stronger algorithms, while FolkEgal tended to not perform well against weaker algorithms.  However, once these weaker algorithms were eliminated, it performed quite well.  {\sc mega-S++} performed relatively well in each grouping.

\begin{table}[t]
\begin{center}
{\scriptsize
\setlength{\tabcolsep}{4pt}
\begin{tabular}[b]{l|ccc|ccc|ccc} \hline
 & \multicolumn{3}{c|}{\bf Microgrid Scenario}  & \multicolumn{3}{c|}{\bf SGPD}  & \multicolumn{3}{c}{\bf Block Game} \\ 
 {\bf Algorithm} & {\bf Self} & {\bf Top} & {\bf Low}  & {\bf Self} & {\bf Top} & {\bf Low}  & {\bf Self} & {\bf Top} & {\bf Low} \\ \hline
{\sc mega-S++} & 32.2 & 26.9 & 20.9 & 15.8 & 11.3 & 11.6 & 23.3 & 17.2 & 28.0 \\
{\sc CFR} & 17.6 & 23.4 & 20.1 & 7.0 & ~~8.4 & 11.4 & 18.0 & 17.0 & 24.4 \\
{\sc FolkEgal} & 33.3 & 24.1 & 12.8 & 16.0 & 10.3 & ~~8.7 & 24.9 & 15.3 & 14.6 \\ \hline
\end{tabular}}
\end{center} 
	\caption{Average payoffs over 500 rounds in self play, against Top-5 competition, and against Low-5 competition.}
	\label{tab:breakdown}
\end{table}

\begin{table}[t]
\begin{center}
{\scriptsize
\setlength{\tabcolsep}{4pt}
\begin{tabular}[b]{c|ccc|ccc|ccc} \hline
	& \multicolumn{3}{c|}{\bf Microgrid Scenario} & \multicolumn{3}{c|}{\bf SGPD} & \multicolumn{3}{c}{\bf Block Game}\\ 
	{\bf Algorithm} & {\bf Init} & {\bf Run} & {\bf Total} & {\bf Init} & {\bf Run} & {\bf Total} & {\bf Init} & {\bf Run} & {\bf Total}\\ \hline 
	{\sc mega-S++} & 9.3 & 0.9 & 10.2 & 131.5 & ~~58.3 & 189.7 & 0.8 & ~~0.9 & ~~1.7 \\
	CFR & 0.1 & 0.8 & ~~1.0 & ~~19.0 & 911.0 & 930.0 & 7.1 & 58.8 & 65.9 \\
	{MBRL} & {0.1} & {1.7} & {~~1.8} & {~~11.7} & {~~79.7} & {~~91.4} & {0.2} & {~~1.1} & {~~1.3} \\ \hline
\end{tabular}}
\end{center}
	\caption{Initialize time, runtime, and total time to complete a 500-round game in self play (averaged over 25 trials).  All times are in seconds.  Simulations were run on a 2.6 GHz Intel Core i7 processor with 16 GB 1600 MHZ DDR3 memory.}
	\label{tab:comptime}
\end{table}

\subsection{Computation Time}
To create the set $\Phi_i$, {\sc mega} solves a constant number of MDPs {\em a priori}, each of which can be done in polynomial time using linear programming.  To create the leader strategies, {\sc mega} computes its maximin and attack strategies, and solves MDP($\omega$) for various $\omega$.  Only minimal computations are required to maintain leader strategies thereafter (to update guilt, etc.).  On the other hand, both MBRL and Bouncer (when selected) require an MDP to be solved after every round.  The other followers utilize computations performed in the creation of the leader experts.  Thus, $\Phi_i$ can be computed and maintained in polynomial time.

In practice, the computation times for creating and maintaining $\Phi_i$ are quite reasonable.  Table~\ref{tab:comptime} shows initialization- and run-times for {\sc mega-S++}, {\sc MBRL}, and {\sc CFR} in 500-round games.   Most of the computation time taken by {\sc mega-S++} was during initialization.  In fact, {\sc mega-S++} typically had shorter runtimes than both {\sc MBRL} and {\sc CFR}.  {\sc CFR} tended to have the longest total execution time in these RSGs.  

We caution that these results were obtained from code that was not heavily optimized.  For example, much of the time required to initialize {\sc mega-S++} in the Microgrid Scenario (and about half the initialization time in the SGPD) was spent in an unoptimized routine for computing minimax.

\section{Conclusion}

Our goal is to identify how to quickly learn effective strategies in general-sum RSGs played against arbitrary associates.  In this paper, we proposed {\sc mega}, a meta-gaming technique designed to reduce the strategy space of RSGs.  {\sc mega} maintains important attributes of the original strategy space, including security, best response, and (some) NEs.  As such, learning algorithms that operate on this reduced strategy space can quickly learn strategies that are effective in the original RSG.  In this way, {\sc mega} facilitates fast and robust learning in general-sum RSGs played against arbitrary associates.  

{\sc mega} differs from previous game-abstraction methods \cite{Gilpin2006,Schnizlein2009,Ganzfried2012,SandholmSingh2012}.  Whereas previous methods seek to reduce the number of states and actions in the game (to make equilibrium computation feasible), {\sc mega} computes a handful of high-level strategies of the game.  These strategy-reduction methods can work in parallel.  In particular, for games with large state spaces, traditional game-abstraction methods can be used to help solve the MDPs that {\sc mega} must solve when computing its experts.

\section*{Acknowledgments}
This work was funded under the Cooperative Agreement between the Masdar Institute of Science and Technology, Abu Dhabi, UAE and the Massachusetts Institute of Technology, Cambridge, MA, USA, Reference Number 02/MI/MIT/CP/11/07633/GEN/G/00.

\appendix

\section{MBRL}
\label{app:mbrl}

The version of model-based reinforcement learning (MBRL) we used models the associate's strategy using the fictitious-play assessment \cite{FictitiousPlay} conditioned on the stage $s \in S$.  That is, in round $t$, player $i$ estimates that $-i$ plays action $a_{-i}$ in stage $s$ with probability $\gamma_{-i}^t(s,a_{-i}) = \frac{\kappa_{-i}^t(s,a_{-i})}{\sum_{b \in A_{-i}(s)} \kappa_{-i}^t(s,b)}$, where $\kappa_{-i}^t(s,a)$ is the number of times that $-i$ has taken action $a$ in $s$ up to round $t$.  For all $a$ and $s$, MBRL computes the value for taking action $a$ in $s$ by solving 
{\scriptsize
\[
Q_i(s,a) = \sum_{b \in A_{-i}(s)} \gamma_{-i}^t(s,b) \left[ r_i(s, ab) + \sum_{s' \in S} P_M(s, ab, s') V_i(s') \right],
\]}where $ab$ is the joint action when $i$ plays $a$ and $-i$ plays $b$ and $V_i(s) = \max_{a \in A_i(s)} Q_i(s,a)$.
We used $\epsilon$-greedy exploration.

{\small
\bibliographystyle{named}
\bibliography{refs.bib}}

\begin{thebibliography}{}

\bibitem[\protect\citeauthoryear{Arora \bgroup \em et al.\egroup
  }{2012}]{Arora2012}
R.~Arora, O.~Dekel, and A.~Tewari.
\newblock Online bandit learning against an adaptive adversary: from regret to
  policy regret.
\newblock In {\em ICML}, pages 1503--1510, 2012.

\bibitem[\protect\citeauthoryear{Auer \bgroup \em et al.\egroup
  }{1995}]{auer95gambling}
P.~Auer, N.~Cesa-Bianchi, Y.~Freund, and R.~E. Schapire.
\newblock Gambling in a rigged casino: the adversarial multi-armed bandit
  problem.
\newblock In {\em Proc. of the 36th Symp. on the Foundations of CS}, pages
  322--331, 1995.

\bibitem[\protect\citeauthoryear{Auer \bgroup \em et al.\egroup }{2002}]{UCB}
P.~Auer, N.~Cesa-Bianchi, and P.~Fischer.
\newblock Finite-time analysis of the multi-armed bandit problem.
\newblock {\em Machine Learning}, 47:235--256, 2002.

\bibitem[\protect\citeauthoryear{Bowling and Veloso}{2002}]{WoLF}
M.~Bowling and M.~Veloso.
\newblock Multiagent learning using a variable learning rate.
\newblock {\em Artificial Intelligence}, 136(2):215--250, 2002.

\bibitem[\protect\citeauthoryear{Bowling}{2004}]{GIGA-WoLF}
M.~Bowling.
\newblock Convergence and no-regret in multiagent learning.
\newblock In {\em NIPS}, pages 209--216, 2004.

\bibitem[\protect\citeauthoryear{Cesa-Bianchi \bgroup \em et al.\egroup
  }{2013}]{CesaBianchi2013}
N.~Cesa-Bianchi, O.~Dekel, and O.~Shamir.
\newblock Online learning with switching costs and other adaptive adversaries.
\newblock In {\em NIPS}, pages 1160--1168, 2013.

\bibitem[\protect\citeauthoryear{Claus and Boutilier}{1998}]{ClausBoutilier}
C.~Claus and C.~Boutilier.
\newblock The dynamics of reinforcement learning in cooperative multiagent
  systems.
\newblock In {\em AAAI}, pages 746--752, 1998.

\bibitem[\protect\citeauthoryear{Crandall}{2012}]{Pepper}
J.~W. Crandall.
\newblock Just add {P}epper: {e}xtending learning algorithms for repeated
  matrix games to repeated markov games.
\newblock In {\em AAMAS}, pages 399--406, 2012.

\bibitem[\protect\citeauthoryear{Crandall}{2014}]{CrandallJAIR2014}
J.~W. Crandall.
\newblock Towards minimizing disappointment in repeated games.
\newblock {\em Journal of Artificial Intelligence Research}, 49:111--142, 2014.

\bibitem[\protect\citeauthoryear{{de~Cote} and Littman}{2008}]{FolkEgal}
E.~{de~Cote} and M.~L. Littman.
\newblock A polynomial-time {Nash} equilibrium algorithm for repeated
  stochastic games.
\newblock In {\em UAI}, pages 419--426, 2008.

\bibitem[\protect\citeauthoryear{{de Farias} and Megiddo}{2004}]{exextrade}
D.~{de Farias} and N.~Megiddo.
\newblock Exploration-exploitation tradeoffs for expert algorithms in reactive
  environments.
\newblock In {\em NIPS}, pages 409--416, 2004.

\bibitem[\protect\citeauthoryear{Dietterich}{2000}]{EnsembleMethods}
T.~G. Dietterich.
\newblock Ensemble methods in machine learning.
\newblock In {\em Multiple Classifier Systems}, {\em LN in CS}, vol 1857, pp.
  1-15. Springer Berlin Heidelberg, 2000.

\bibitem[\protect\citeauthoryear{Elidrisi \bgroup \em et al.\egroup
  }{2014}]{Elidrisi2014}
M.~Elidrisi, N.~Johnson, M.~Gini, and J.~W. Crandall.
\newblock Fast adaptive learning in repeated stochastic games by game
  abstraction.
\newblock In {\em AAMAS}, 2014.

\bibitem[\protect\citeauthoryear{Fudenberg and Levine}{1998}]{FictitiousPlay}
D.~Fudenberg and D.~Levine.
\newblock {\em The Theory of Learning in Games}.
\newblock The MIT Press, 1998.

\bibitem[\protect\citeauthoryear{Ganzfried \bgroup \em et al.\egroup
  }{2012}]{Ganzfried2012}
S.~Ganzfried, T.~Sandholm, and K.~Waugh.
\newblock Strategy purification and thresholding: Effective non-equilibriuam
  approaches for playing large games.
\newblock In {\em AAMAS}, 2012.

\bibitem[\protect\citeauthoryear{Gilpin and Sandholm}{2006}]{Gilpin2006}
A.~Gilpin and T.~Sandholm.
\newblock A competitive {Texas Hold'em} poker player via automated abstraction
  and real-time equilibrium computation.
\newblock In {\em AAAI}, 2006.

\bibitem[\protect\citeauthoryear{Gintis}{2000}]{GintisGameTheory}
Herbert Gintis.
\newblock {\em Game Theory Evolving: A Problem-Centered Introduction to
  Modeling Strategic Behavior}.
\newblock Princeton University Press, 2000.

\bibitem[\protect\citeauthoryear{Greenwald and Hall}{2003}]{CEQ}
A.~Greenwald and K.~Hall.
\newblock Correlated {Q}-learning.
\newblock In {\em ICML}, pages 242--249, 2003.

\bibitem[\protect\citeauthoryear{Hu and Wellman}{1998}]{HuandWellman}
J.~Hu and M.~P. Wellman.
\newblock Multiagent reinforcement learning: Theoretical framework and an
  algorithm.
\newblock In {\em ICML}, pages 242--250, 1998.

\bibitem[\protect\citeauthoryear{Johanson \bgroup \em et al.\egroup
  }{2012}]{Johansonetal2012}
M.~Johanson, N.~Bard, M.~Lanctot, R.~Gibson, and M.~Bowling.
\newblock Evaluating state-space abstractions in extensive-form games.
\newblock In {\em AAMAS}, pages 837--846, 2012.

\bibitem[\protect\citeauthoryear{Littman and Stone}{2001}]{LittmanLeaderAlgs}
M.~L. Littman and P.~Stone.
\newblock Leading best-response strategies in repeated games.
\newblock In {\em IJCAI workshop on Economic Agents, Models, and Mechanisms},
  Seattle, WA, 2001.

\bibitem[\protect\citeauthoryear{Littman \bgroup \em et al.\egroup
  }{1995}]{LittmanMDP}
M.~L. Littman, T.~L. Dean, and L.~P. Kaelbling.
\newblock On the complexity of solving markov decision problems.
\newblock In {\em UAI}, 1995.

\bibitem[\protect\citeauthoryear{Littman}{1994}]{minimaxQ}
M.~L. Littman.
\newblock Markov games as a framework for multi-agent reinforcement learning.
\newblock In {\em ICML}, pages 157--163, 1994.

\bibitem[\protect\citeauthoryear{Littman}{2001}]{friendorfoe}
M.~L. Littman.
\newblock Friend-or-foe: {Q}-learning in general-sum games.
\newblock In {\em ICML}, pages 322--328, 2001.

\bibitem[\protect\citeauthoryear{Oudah \bgroup \em et al.\egroup
  }{2015}]{Oudah2015}
M.~Oudah, V.~Babushkin, T.~Chenlinangjia, and J.~W. Crandall.
\newblock Learning to interact with a human partner.
\newblock In {\em HRI}, 2015.

\bibitem[\protect\citeauthoryear{Papadimitriou and
  Tsitsiklis}{1987}]{PapadimitriouTsitsiklis1987}
C.~H. Papadimitriou and J.~N. Tsitsiklis.
\newblock The complexity of {Markov} chain decision processes.
\newblock {\em Mathematics of Operations Research}, 12(2):441--450, 1987.

\bibitem[\protect\citeauthoryear{Powers and Shoham}{2005}]{PowersIJCAI2005}
R.~Powers and Y.~Shoham.
\newblock Learning against opponents with bounded memory.
\newblock In {\em IJCAI}, pages 817--822, 2005.

\bibitem[\protect\citeauthoryear{Rummery and Niranjan}{1994}]{Sarsa1994}
G.~A. Rummery and M.~Niranjan.
\newblock On-line {Q-learning} using connectionist sytems.
\newblock Technical Report CUED/F-INFENG-TR 166, Cambridge University, UK,
  1994.

\bibitem[\protect\citeauthoryear{Sandholm and Singh}{2012}]{SandholmSingh2012}
T.~Sandholm and S.~Singh.
\newblock Lossy stochastic game abstraction with bounds.
\newblock In {\em EC}, pages 880--897, 2012.

\bibitem[\protect\citeauthoryear{Schnizlein \bgroup \em et al.\egroup
  }{2009}]{Schnizlein2009}
D.~Schnizlein, M.~Bowling, and D~Szafron.
\newblock Probabilistic state translation in extensive games with large action
  sets.
\newblock In {\em IJCAI}, 2009.

\bibitem[\protect\citeauthoryear{Taylor and Jonker}{1978}]{ReplicatorDynamic}
P.~D. Taylor and L.~Jonker.
\newblock Evolutionarily stable strategies and game dynamics.
\newblock {\em Mathematical Biosciences}, 40:145--156, 1978.

\bibitem[\protect\citeauthoryear{Zinkevich \bgroup \em et al.\egroup
  }{2007}]{Zinkevich2007}
M.\ Zinkevich, M.\ Bowling, M.\ Johanson, and C.\ Piccione.
\newblock Regret minimization in games with incomplete information.
\newblock In {\em NIPS}, 2007.

\end{thebibliography}

\end{document}